\documentclass[10pt,journal,a4paper]{IEEEtran}
\makeatletter
\def\ps@headings{%
\def\@oddhead{\mbox{}\scriptsize\rightmark \hfil \thepage}%
\def\@evenhead{\scriptsize\thepage \hfil \leftmark\mbox{}}%
\def\@oddfoot{}%
\def\@evenfoot{}}
\makeatother
\usepackage{multirow}
\usepackage{amssymb}
\usepackage{amsmath}
\usepackage{amsfonts}
\usepackage{graphicx}
\usepackage{subfigure}
\usepackage{cite}
\usepackage{enumerate}
\usepackage{url}
\usepackage{bm}
\usepackage{clrscode}
\usepackage{subfigure}
\usepackage{amsthm}

\begin{document}

\title{Beyond D2D: Full Dimension UAV-to-Everything Communications in 6G}


\author{\IEEEauthorblockN{
		{Shuhang Zhang}, \IEEEmembership{Student Member, IEEE},
		{Hongliang Zhang}, \IEEEmembership{Member, IEEE},\\
        and {Lingyang Song}, \IEEEmembership{Fellow, IEEE}\\
\thanks{S. Zhang and L. Song are with Department of Electronics, Peking University, Beijing, 100871 China (email: \{shuhangzhang, lingyang.song\}@pku.edu.cn).}
\thanks{H. Zhang is with Department of Electronics, Peking University, Beijing, 100871 China, and also with Electrical and Computer Engineering Department, University of Houston, Houston, TX 77004 USA (email: hongliang.zhang92@gmail.com).}
}}
\maketitle
\setlength{\abovecaptionskip}{0pt}
\setlength{\belowcaptionskip}{-10pt}
\begin{abstract}
In this paper, we consider an Internet of unmanned aerial vehicles~(UAVs) over cellular networks, where UAVs work as aerial users to collect various sensory data, and send the collected data to their transmission destinations over cellular links. Unlike the terrestrial users in the conventional cellular networks, different UAVs have various communication requirements due to their sensing applications, and a more flexible communication framework is in demand. To tackle this problem, we propose a UAV-to-Everything~(U2X) networking, which enables the UAVs to adjust their communication modes full dimensionally according to the requirements of their sensing applications. In this article, we first introduce the concept of U2X communications, and elaborate on its three communication modes. Afterwards, we discuss the key techniques of the U2X communications, including joint sensing and transmission protocol, UAV trajectory design, and radio resource management. A reinforcement learning-based mathematical framework for U2X communications is then proposed. Finally, the extensions of the U2X communications are presented.
\end{abstract}
\section{Introduction}
While the fifth generation (5G) communication has been expected to be utilized in various industrial fields, it is considered that a wider diffusion will be required as a type of further development in the future sixth generation~(6G) era~\cite{DOCOMO}. An extremely large number of Internet of Things~(IoT) devices that collect images and sensing information of the real world are expected to spread further in the 6G networks, and an extremely large number of connections that are approximately 10 fold (10 million devices per square km) more than the 5G requirements are needed~\cite{ZXMXDLKF2019}. In addition to the approach of connecting a large number of IoT devices to a network, the IoT devices are expected to evolve to have functions for detecting the real world and collecting various sensory data intelligently~\cite{ZFWDWW2019}.

\begin{figure}[t]
\centerline{\includegraphics[width=3.5in]{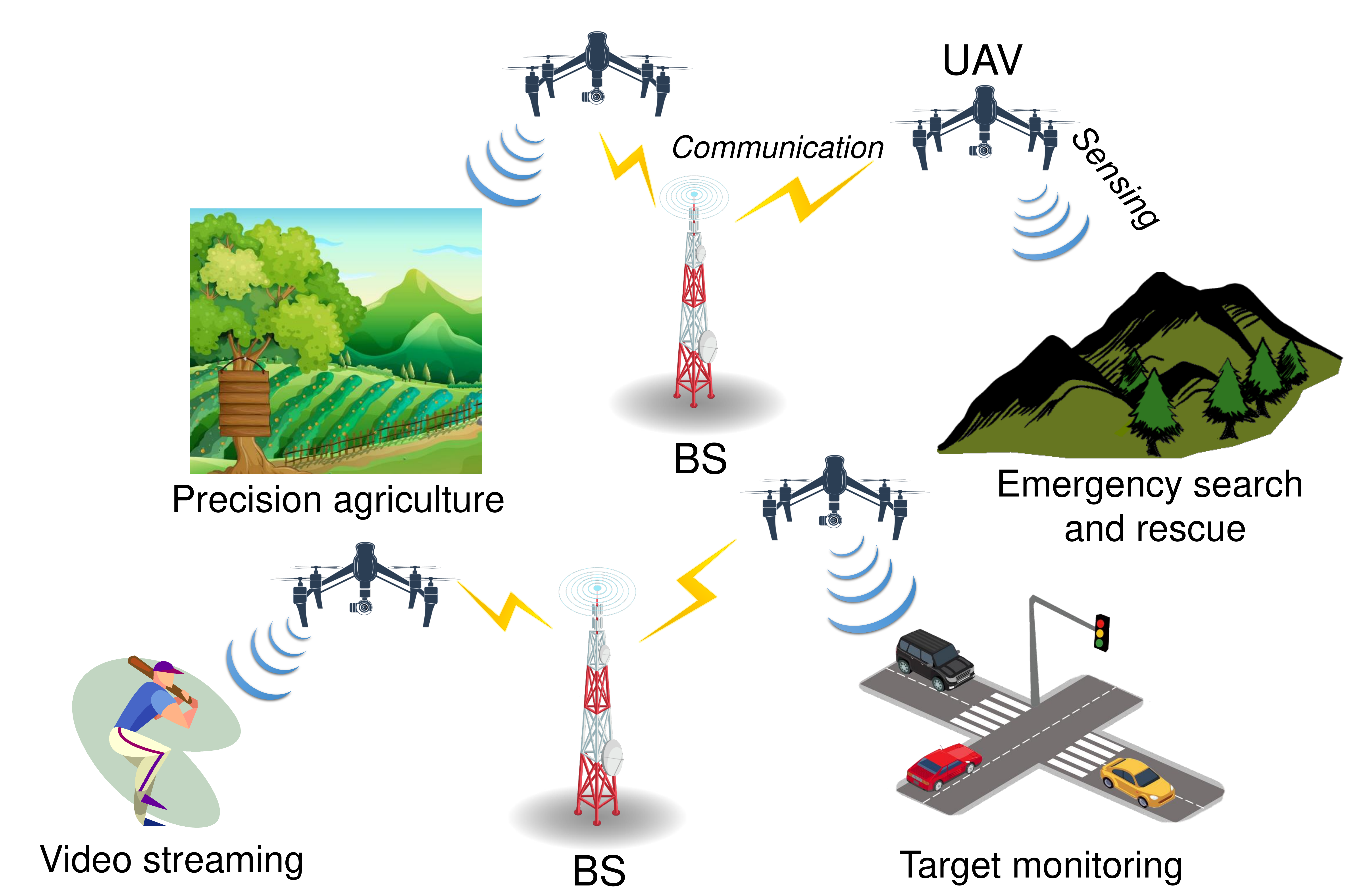}}
\caption{Illustration for the cellular Internet of UAVs applications.}
\label{application}
\end{figure}

Unmanned aerial vehicles~(UAVs), as a type of aerial IoT devices, have attracted the interests from academic and industry for real-time sensing applications, owing to the advantages of high mobility and large service coverage~\cite{FTNKM2016},~\cite{ZZHZXWWZ2019}. According to a Business Insider Intelligence's report, more than 29 million UAVs are expected to be put into use in 2021, where the UAVs for sensing applications take up a major part of the market share~\cite{UAVmarket}. As illustrated in Fig.~\ref{application}, UAVs with onboard sensors are deployed in the cellular networks to provide various sensing services, which is called the cellular Internet of UAVs~\cite{ZSH2019}. In the cellular Internet of UAVs, the sensory data is transmitted to the terrestrial user equipments~(UEs) directly or to a remote sever through the base stations (BSs) according to different applications~\cite{KNKOM2019}. Recently, various sensing applications have been discussed in the cellular Internet of UAVs. In~\cite{KGM2017}, a UAV was utilized to sense the air quality samples of some targets, and transmit the data to the requiring UEs. In~\cite{AMCG2017}, the authors built a precision agriculture platform in which UAVs sense the condition of the crops with cameras, and upload the data to a central server, which processes the sensory data and takes actions to the crops correspondingly.

The cellular Internet of UAVs multiplexes the spectrum resources and infrastructure of the terrestrial cellular UEs, and consumes the communication services supported by the powerful hardware foundation in the 6G era~\cite{ZZHBS2017}. However, due to the diverse requirements for UAV communications, the UAVs are incapable to achieve high data rate by applying the terrestrial cellular networks directly~\cite{XZLNWG2020}. Unlike the conventional terrestrial UEs that can obtain high transmission rate by communicating with a nearby BS, the sensing UAVs have full dimension transmission destinations, including BS, terrestrial UEs, and other adjacent UAVs~\cite{TFKOM2018}. Therefore, a more flexible framework is necessary for the UAVs, in which they have the degree of freedom to choose the optimal transmission links in full dimensions~\cite{ZZDS2019}. Moreover, different from the terrestrial UEs, the UAVs are likely to move to the cell edge for the sake of sensing performance. On this condition, it is difficult to guarantee high transmission rate with the conventional cellular framework, where all the sensory data are required to be uploaded to the BS over cellular links.

To tackle the above challenges, we propose the concept of full dimension UAV-to-Everything (U2X) communications, which contains three different modes, i.e., UAV-to-Network (U2N), UAV-to-UAV (U2U), and UAV-to-Device (U2D) communications. For U2N communications, the sensory data of each UAV is sent to the BS via cellular links. The BS can then transmit the data to the destination nodes in the backbone network, and thereby providing high reliable data transmissions. For U2U communications, two UAVs in proximity can set up a direct link bypassing the BS, which enables the UAVs to perform cooperative sensing and transmission~\cite{ZSHP2019}. The data rates for U2U communications can be further improved by exploiting proximity gain and underlay nature. For U2D communications, the UAV transmits the sensory data to its destination node directly, with which the transmission can be completed within single-hop, thereby reducing the transmission latency. The U2X communications enable the UAVs to adopt different transmission modes according to the specific requirements of their corresponding sensing applications, and provide a feasible architecture for the utilization of UAV sensing in the 6G network~\cite{GLTKA2020,KMTKL2020}.

Note that the full dimension U2X communications in the cellular Internet of UAVs is different from the existing works on the cellular vehicular to everything~(V2X) for the following three reasons. First, the U2X communications are mostly utilized for data sensing and transmission, while the V2X communications are designed for improving road safety and traffic efficiency in priority~\cite{CHSPFZZ2017}. Therefore, the optimization target of U2X and V2X communications are different. Second, the UAVs can design their trajectories in a three-dimensional space to complete the sensing applications, while the vehicles can only move along the given roads, and the destinations of the vehicles are set by the drivers in advance~\cite{DSLH2017}. As a result, the trajectory design for the cellular Internet of UAVs is a new concept which has not been mentioned in the cellular V2X communications. Third, the UAVs perform UAV-to-ground communications to transmit the sensory data to the BS and terrestrial UEs, and the channel model for U2X and V2X communications are totally different~\cite{JZG2019}. In summary, the service requirements, key challenges, and physical models of the U2X communications are significantly different from the existing V2X communications, and need to be studied independently.

In this article, we first introduce the basics of U2X communications, including the model of the cellular Internet of UAVs and the three communication modes. Then, we present the key techniques for the U2X communications, consisting of joint sensing and transmission protocol, UAV trajectory design, and radio resource management.
\begin{enumerate}
\item \textbf{Joint Sensing and Transmission Protocol:} The sensing and transmission processes are coupled together, and need to be designed jointly. A joint U2X sensing and transmission protocol considering the decentralized communication mode selection and centralized radio resource management is required.
\item \textbf{UAV Trajectory Design:} The trajectory of the UAVs should be designed based on the locations of the sensing targets and the transmission destinations. Moreover, a decentralized trajectory design algorithm is required since it is very difficult for the BS to obtain the realtime channel state information~(CSI) of the highly mobile UAVs.
\item \textbf{Radio Resource Management:} To further improve the spectrum efficiency, the U2U and U2D communication links can share the spectrum with the U2N and cellular ones. Therefore, the radio resource management, including the subchannel allocation and the power control, is necessary to reduce the co-channel interference.
\end{enumerate}
Afterwards, we propose a reinforcement learning (RL) based framework to maximize the probability for successful sensing and transmission with U2X communications. The UAV trajectory design and radio resource management are solved, and the performance of the proposed framework is evaluated. In addition, we also introduce some extensive scenarios of the U2X communications, and discuss the corresponding open problems and potential solutions, as illustrated below:
\begin{enumerate}
\item \textbf{UAV Cooperation with U2X Communications:} Some of the sensing applications require multiple UAVs to perform sensing and transmission for a task cooperatively. To reduce the cost of power and spectrum resources, the UAVs converge their sensory data to one of the UAVs, which transmits all the data to the destination uniformly.
\item \textbf{Multi-access Edge Computing (MEC) with U2X Communications:} The UAVs that possess computing capabilities offer data processing to the cell edge UEs with limited local computing capabilities. The MEC reduces the computation workload of the BS, and improves the quality of service (QoS) of the transmissions to the cell edge UEs.
\item \textbf{Non-Orthogonal Multiple Access (NOMA) for U2X Communications:} NOMA can be utilized in the U2X communications for the underlay and overlay mode transmissions. In this way, massive U2X communication links can be supported with limited spectrum resource, and the network throughput can be further improved.
\end{enumerate}

The main contributions of this paper can be summarized as below.
\begin{enumerate}
\item We propose the concept of U2X communications in the cellular Internet of UAVs, which contains three communication modes: U2N, U2U, and U2D modes. The system model of UAV sensing and UAV transmission and the characteristics of the three communication modes are elaborated on.
\item We study the key techniques of U2X communications, including joint sensing and transmission protocol, trajectory design, and radio resource management. A RL framework based algorithm is utilized to estimate the performance of the U2X communications.
\item We give the extensions of the U2X communications, such as UAV cooperation, MEC, and NOMA, and raise some open problems correspondingly. The related works and potential solutions to the open problems are also discussed.
\end{enumerate}

The rest of this article is organized as follows. We first introduce the basics of U2X communications in Section~\ref{Basics Sec}. Afterwards, we discuss the key techniques with U2X communications in Section~\ref{technique-sec}, including the main challenges, possible solutions, and performance evaluations. In Section~\ref{math section}, we formulate a UAV sensing
and transmission optimization problem, and solve it in a RL based framework. In Section~\ref{application sec}, we propose three extensive scenarios with U2X communications. Finally, we draw the conclusions in Section~\ref{conclusion sec}.
\section{Basics of U2X Communications}\label{Basics Sec}
In this section, we first introduce the cellular Internet of UAVs, and then illustrate three communication modes utilized in this network.
\subsection{The Cellular Internet of UAVs}\label{architecture sec}
\begin{figure}[t]
\centerline{\includegraphics[width=3.5in]{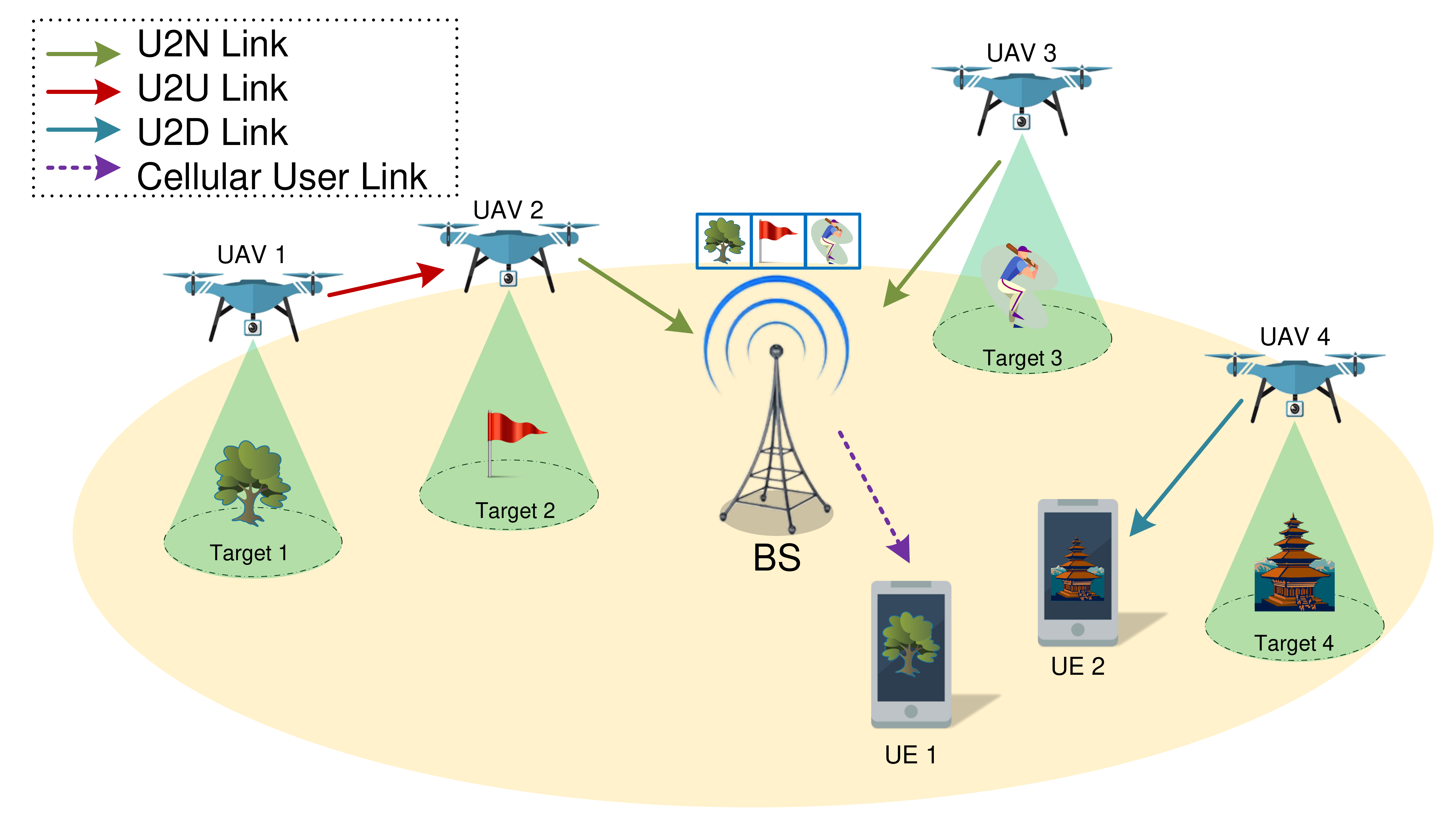}}
\caption{The cellular Internet of UAVs with U2X communications.}
\label{systemmodel}
\end{figure}
Fig.~\ref{systemmodel} shows a cellular Internet of UAVs, which consists of one BS, multiple UAVs, and multiple terrestrial UEs. The UAVs collect data from sensing targets, and transmit the collected data to the BS or the terrestrial UEs to support various applications. The sensing tasks are performed in two steps: UAV sensing and UAV transmission. We define \emph{cycle} as a time unit for the UAV sensing and UAV transmission. In each cycle, the UAVs first sense the required data, and then transmit the sensory data to the corresponding destinations. Iterations of UAV sensing and UAV transmission are performed in a sequence of cycles to support the sensing applications, and the UAVs move along the designed trajectory simultaneously in each cycle. The detailed protocol for UAV sensing and UAV transmission will be introduced in Section~\ref{protocol-sec}.

\subsubsection{UAV Sensing}
The UAVs collect sensory data for the required sensing tasks\footnote{The sensing task assignment can be performed in advance as proposed in~\cite{ZZDS2019'}, and thus, each UAV will perform one task at any time.}. To complete a sensing task, each UAV needs to collect the sensory data from a specific location, which is defined as the \emph{sensing target} of this task. To evaluate the sensing performance, we adopt a probabilistic model related to the distance between the location of the UAV and the sensing target~\cite{ZYZS2019}. The success of UAV sensing is a necessary condition for the completion of the sensing task.
\subsubsection{UAV Transmission}
After UAV sensing, the collected data will be transmitted to the corresponding BS or UE, which is defined as the \emph{transmission destination}. For each UAV, the transmission channel model and frequency band are regulated by the 3GPP~\cite{3GPPR12}. To satisfy the transmission requirements raised by various applications, we provide three communication modes for the UAVs, namely U2N communications, U2U communications, and U2D communications. The communication mode for each link is selected by the UAV according to its location and the communication requirements of the sensing application. The descriptions of these three communication modes will be elaborated on in Section~\ref{U2X mode sec}.
\subsection{U2X Communication Modes}\label{U2X mode sec}
In this part, we discuss the full dimension U2X communications, which contain three possible communication modes for sensory data transmission in this network.

\subsubsection{UAV-to-Network Mode}\label{U2N sec}
In U2N mode, a UAV uploads its collected data to the BS directly overlaying the cellular uplink spectrum resource\footnote{The U2X communication overlaying or underlaying the cellular network can be implemented as introduced in some existed works~\cite{ZLS2017}. Techniques such as cognitive radio~\cite{LSHL2015} can be utilized for spectrum sharing in the cellular Internet of UAVs to reduce the interference.}. U2N communication provides high data rate and low latency transmission when the UAV is close to the BS. It also enables multiple UAVs to communicate with the BS over different subchannels simultaneously, which reduces the potential transmission interference. U2N communication is essential for the applications that require the BS to collect massive data, such as crowdsourcing. Furthermore, the existing infrastructure guarantees high downlink rate for the terrestrial cellular UEs, and enable the UEs to access highly reliable sensory data collected by the BS from the UAVs.
\subsubsection{UAV-to-UAV Mode}
In U2U mode, a UAV communicates with another UAV underlaying the cellular and U2N communications, which reduces the latency and provides high QoSs for the communications between adjacent UAVs. U2U communications also allow a UAV to broadcast data to every direction that has adjacent UAVs, thus providing physical mechanism for efficient information diffusion. Moreover, the underlaying transmission improves the spectrum efficiency of the network, which is necessary for the network with massive UEs.
\subsubsection{UAV-to-Device Mode}
In U2D mode, a UAV communicates with destination UE directly underlaying the cellular and U2N communications. U2D communications allow the UEs to receive the data from a UAV directly bypassing the BS. U2D communication is especially efficient when a UAV is close to its transmission destination UE. It not only saves the spectrum and energy resources for UAV-to-BS and BS-to-UE links, but also reduces the transmission latency. It is promising for low latency applications without a huge amount of computation, such as live video streaming.

\section{Key Techniques for U2X Communications}\label{technique-sec}
In this section, we introduce the key techniques that support the U2X communications in the cellular Internet of UAVs. To support the U2X communications, it is necessary to discuss the key techniques that solves the following three challenges. First, since UAV sensing and UAV transmission are coupled together, a joint sensing and transmission protocol for U2X communications is required. Second, the trajectories of the highly mobile UAVs should be designed properly to maximize the utilities of the UAVs determined by their sensing applications~\cite{TKNKOM2018}. Third, the UAVs with different transmission modes may share the spectrum resources to improve the network throughput. As a result, the spectrum and power resources should be managed properly.
\subsection{Joint Sensing and Transmission Protocol}\label{protocol-sec}
In this part, we propose a joint sensing and transmission protocol to coordinate multiple UAVs performing sensing tasks simultaneously in different transmission modes. The protocol for U2X communications is more complicated than the protocols in other previous works~\cite{ZZDS2019,ZSHP2019} for the following reason. On the one hand, it is difficult for the BS to obtain the realtime CSI of all the UAVs, and the communication mode selection and UAV trajectory need to be performed by the UAVs in a decentralized manner. On the other hand, the spectrum resources of the UAVs should be managed by the BS in a centralized manner since the UAVs share the spectrum resources of the cellular UEs. Therefore, this protocol should consider the designs performed by both the UAVs and the BS.

\begin{figure}[t]
\centerline{\includegraphics[width=3in]{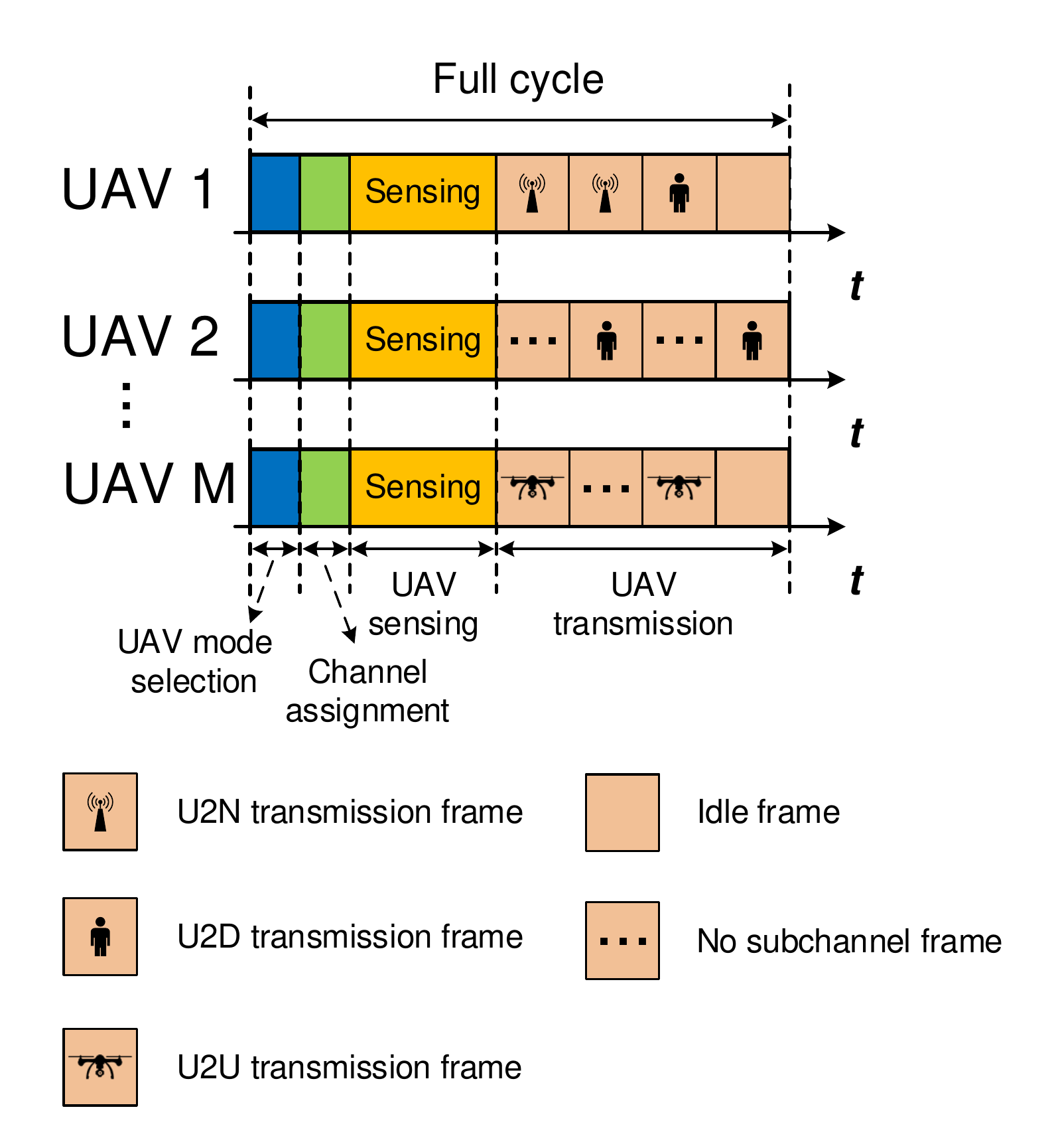}}
\caption{The joint sensing and transmission protocol.}
\label{protocol}
\end{figure}

We assume that the UAVs perform sensing and transmission in a synchronized iterative manner, which is characterized by \emph{cycles}. In Fig.~\ref{protocol}, we illustrate a cycle of the proposed protocol, which mostly consists of four parts: UAV mode selection, channel assignment, UAV sensing, and UAV transmission.

\begin{itemize}
\item \textbf{UAV mode selection:} The transmission mode of each communication link is selected by the UAVs independently. Although a UAV can have multiple transmission destinations, only one transmission mode can be adopted by each UAV-to-destination link, and the transmission modes to different destinations can be different, i.e. a UAV can adopt multiple transmission modes if it has multiple transmission destinations. Each UAV sets a QoS threshold for its transmissions according to its sensing application, and adopts the transmission mode to satisfy the QoS threshold for each link\footnote{The UAVs only consider the large scale channel fading and estimate the average transmission QoS, since it is difficult to obtain the real-time small scale fading.}. When multiple transmission modes satisfy the transmission QoS threshold, the UAV selects the mode with the highest data rate. If none of these transmission modes satisfy the QoS threshold, the UAV moves along the designed trajectory until the threshold can be satisfied by at least one transmission mode.

\item \textbf{Channel assignment:} The BS first obtains the locations of the UAVs and that of their sensing targets, and then broadcasts the locations of all the UAVs. By this means, each UAV can obtain the locations of other UAVs, and then decides its trajectory in the next cycle, which will be introduced in Section~\ref{Trajectory Sec}. Afterwards, the BS performs the subchannel allocation and power control for the UAVs and terrestrial UEs. The results of the radio resource management are then sent to the UAVs and terrestrial UEs over control channel.

\item \textbf{UAV sensing:} Each UAV senses its target continuously for a period of time to ensure the sensing quality. The requirements for UAV sensing is determined by the specific type of sensing application.
\item \textbf{UAV transmission:} The UAVs transmit the sensory data to their transmission destinations. Each cycle contains multiple transmission frames, and a UAV can adopt only one transmission mode in each frame. To be specific, there are five possible situations for each UAV in the transmission frame:
\begin{enumerate}
\item \emph{U2N transmission frame:} The UAV accesses to a subchannel, and transmits the sensory data in the U2N mode;
\item \emph{U2D transmission frame:} The UAV accesses to a subchannel, and transmits the sensory data in the U2D mode;
\item \emph{U2U transmission frame:} The UAV accesses to a subchannel, and transmits the sensory data in the U2U mode;
\item \emph{Idle frame:} The data transmission is completed if all the data sensed by the UAV has been transmitted. Then, the UAV will keep idle until the next cycle begins;
\item \emph{No subchannel frame:} No subchannel is allocated to the UAV, and thus, the UAV cannot transmit data in this frame. It will wait for the subchannel access in the following frames.
\end{enumerate}
\end{itemize}
\subsection{UAV Trajectory Design}\label{Trajectory Sec}
The design of UAV trajectory is a key technique in the cellular Internet of UAVs. UAV trajectory determines the location of a UAV in the future cycles, which has a vital impact on the performance of both UAV sensing and UAV transmission. To be specific, when the location of a sensing target is given, the successful sensing probability is determined by the location of the UAV, which is a function of the UAV trajectory. In the meanwhile, given the location of its transmission destination, the received signal-to-noise ratio (SNR) is also a function of the UAV location. Since the UAV sensing and UAV transmission processes are coupled together in each cycle, the design of UAV trajectory should consider both of them simultaneously. However, the performance of UAV sensing and UAV transmission may not be improved concurrently. To be specific, the UAV needs to move toward the sensing target to improve the corresponding successful sensing probability, while it can improve the transmission QoS by approaching the transmission destination. When the sensing target and the transmission destination are on different directions, a trade-off between the successful probability of UAV sensing and UAV transmission should be considered in the design of UAV trajectory. The specific trade-off principle is determined by the sensing and transmission requirements of the corresponding sensing application.

To design the UAV trajectory for joint sensing and transmission, the following challenges should be considered. First, since the UAVs for various sensing applications have different transmission modes and communication requirements, it is especially difficult for the BS to perform joint trajectory design for the UAVs in a centralized manner. Therefore, a decentralized trajectory design algorithm performed by each UAV independently is necessary. Second, for a U2U communication link, the transmission QoS is determined by the trajectories of both the transmission UAV and the receiving UAV. As a result, the trajectory design method should allow a UAV to predict the performance of the others. Third, due to the existence of the co-channel interference, the trajectory design problem is a non-convex problem~\cite{ZZDS2019}, which is difficult to be processed mathematically, and a proper method is required to solve this problem.

To tackle the above challenges, multi-agent deep RL is considered to be a promising method~\cite{TKKL2019,XZNDWW2019}, in which each of the UAVs is considered as an agent that can design its own trajectory independently. A UAV designs its trajectory by continuously interacting with the system environment, including the reacts of the BS and other UAVs, to maximize its own reward determined by the sensing application. The reward of the UAV trajectory is measured by learning the actions and rewards in the former cycles, rather than solving the non-convex problem directly. A specific example of joint sensing and transmission UAV trajectory design in the multi-agent deep RL framework will be introduced in Section~\ref{RL trajectory subsection}.
\subsection{Radio Resource Management}\label{Resource Sec}
In this part, we discuss the radio resource management in this network, i.e., subchannel allocation and power control. Since the UAVs share the spectrum resources of the terrestrial cellular UEs, their radio resources need to be managed by the BS jointly. However, it is very difficult for the BS to obtain the realtime CSIs of the highly mobile UAVs. Therefore, the BS only considers the large scale fading, which is mostly determined by the locations of the UAVs and their transmission targets, for the radio resource management.
\subsubsection{Subchannel Allocation}\label{subcahnnel Sec}
Subchannel allocation is a deterministic factor on the success of data transmission. A UAV can transmit the sensory data to its transmission destinations only when it accesses to at least one subchannel. In the cellular Internet of UAVs, the U2N communications overlay the spectrum resources of the terrestrial UEs, while the U2U and U2D communications underlay the spectrum resources of the terrestrial UEs and U2N communications. A proper subchannel allocation method is necessary to improve the spectrum efficiency and reduce the co-channel interference. Note that the subchannel allocation problem in the cellular Internet of UAVs is different from the conventional D2D network since the U2N and U2D communications utilize the air-to-ground channel model, which cannot be solved by the methods designed for the ground-to-ground subchannel allocation methods. 

When performing subchannel allocation, the BS considers the transmission power of the UAVs as given values. For the U2N mode transmissions that overlay the cellular network, the subchannel allocation problem can be relaxed to a convex optimization problem. The optimal solution to the U2N mode subchannel allocation can be obtained around the solution to the relaxed convex problem. For the U2U and U2D mode communications that underlay the cellular network, the subchannel allocation is non-convex due to the co-channel interference. A matching based iterative algorithm can be utilized to solve this problem~\cite{ZDSL2017}. The set of UAVs and the set of subchannels are considered as two disjoint sets players aiming to maximize their own utilities. The utility of a UAV is determined by the requirement of its sensing application, e.g., the transmission rate or the outage probability, and the utility of a subchannel can be defined as a variable that measures the performance of the network, such as the data throughput over it. In an iteration, each UAV proposes to access to the subchannel that maximizes its utility. For a subchannel that receives a proposal from the UAV, it will judge if the proposal can improve its own utility, and only the proposals that improve the utility of the subchannel will be accepted. The iterative algorithm terminates after convergence, and the convergent result is the subchannel allocation strategy given by the BS.

\subsubsection{Power Control}\label{power Sec}
Beside the subchannel allocation, the UAV transmission power is also a key impact factor on the success of data transmission. Since the data rate of each link is determined by the transmission SNR, a larger UAV transmission power can improve the transmission QoS of the corresponding U2X communication link. However, due to the existence of underlaying U2U and U2D communications, a larger UAV transmission power also leads to a more severe co-channel interference to the other transmission links sharing the same subchannel. Therefore, the transmission power of the U2X communication links over every subchannel need to be designed jointly, to maximize the sum of the utilities of all the transmissions. Note that the utility function of each transmission link can be different, which is determined by the type of the sensing application.

Given the solution to the subchannel allocation and the trajectory of each UAV, the BS jointly designs the transmission power of the UAVs to improve the spectrum efficiency and reduce the potential co-channel interference brought in by the underlay mode U2U and U2D communications. The transmission power over each subchannel can be designed independently since there is no inter-channel interference. If a subchannel is assigned to only one transmission link in a frame, the UAV power control is a linear problem and can be solved directly. However, when a subchannel is assigned to multiple UAVs, the power control problem becomes non-convex due to the existence of co-channel interference. The non-convex problem can be solved with proper mathematical approximations based on difference of convex algorithm~\cite{TKNKOM2018'}.

To be specific, each UAV has a utility index, which is a function of the transmission power, such as its transmission rate. The BS can set a weight coefficient for the utility of each UAV, which can be determined by the transmission SNR requirement and the size of data that needs to be transmitted. We then convert the problem into maximizing the sum of the weighted utility over this subchannel. The weighted utility of a transmission link can be expressed as the difference of two convex functions, which can be solved by successively approximating the feasible set by a sequence of polyhedral convex sets, and turning the problem into a series of convex problems.

\section{Reinforcement Learning Framework for U2X Communications}\label{math section}
In this section, we introduce an example of sensing and transmission design with U2X communications in the cellular Internet of UAVs. An algorithm based on RL framework is proposed to solve the UAV trajectory design and radio resource management problems, and the performance of the proposed algorithm is evaluated by simulations.

\subsection{Problem Description}
We consider a cellular Internet of UAVs as shown in Fig.~\ref{systemmodel}, in which each of the UAVs is required to sense its assigned task, and then transmit the sensory data to the BS or UEs by U2X communications according to the joint sensing and transmission protocol as proposed in Section~\ref{protocol-sec}. In each cycle, we define that the data transmission of a UAV is \emph{valid} if the UAV successfully senses its target as well as transmits the sensory data to its transmission destinations.

As an example, we aim to maximize the average number of valid data transmissions in each cycle by UAV trajectory design and radio resource management. The average number of valid data transmissions can be expressed as the sum of the expectation of the valid data transmissions in each cycle, which equals the sum of the successful sensing probabilities times the successful transmission probabilities. For UAV-to-destination link $i$ in cycle $t$, $P^s_i(t)$ and $P^t_i(t)$ are the successful probability for the sensing and transmission, respectively. We denote the set of variables for UAV trajectory, subchannel allocation, and transmission power by $\textbf{L}$, $\Psi$, and $P$, respectively. The subchannel allocation is denoted by a binary variable $\psi_{i,j}(t)$, where $\psi_{i,j}(t)=1$ if subchannel $j$ is assigned to UAV-to-destination link $i$ in cycle $t$, and $\psi_{i,j}(t)=0$ otherwise. The transmission power of UAV $i$ in cycle $t$ is denoted by $p_i(t)$.

To maximize the average number of valid data transmissions, both UAV sensing and UAV transmission should be considered. Given the location of a sensing target, the successful sensing probability $P^s_i(t)$ is determined by the location of the UAV, which is a function of the UAV trajectory $\textbf{L}$. Given the transmission QoS requirements for each UAV-to-destination link, the successful transmission probability $P^t_i(t)$ can be defined as a function of the subchannel allocation variables $\Psi$, the UAV trajectory $\textbf{L}$, the transmission power $P$, and the amount of data to be transmitted, which varies for different sensing applications~\cite{HZS2019}. The maximization for the average number of valid data transmissions can be formulated as
\begin{subequations} \label{system_optimization}
\begin{align}
\mathop{\min}\limits_{\textbf{L}, \Psi, P}& \sum P^s_i(t) P^t_i(t), \label{system_1}\\
\textbf{\emph{s.t. }}
&\textbf{l}_i(t+1)-\textbf{l}_i(t)\leq v_{max},\label{system_2}\\
&\psi_{i,j}(t)=\{0,1\}, \forall \psi_i(t)\in \Psi,\label{system_3}\\
&p_i(t)\leq p_{max}, \forall p_i(t)\in P.\label{system_4}
\end{align}
\end{subequations}
The velocity constraint for the UAV trajectory design is given in~(\ref{system_2}).~(\ref{system_3}) is a binary subchannel allocation constraint, and~(\ref{system_4}) is the maximum transmission power constraint.
In the following, we propose a RL based algorithm to solve the UAV trajectory design and radio resource management.
\subsection{Brief Introduction to Reinforcement Learning}
In this part, we introduce the concept of RL briefly. A RL framework algorithm will be introduced in Section~\ref{RL trajectory subsection} to solve problem (\ref{system_optimization}).
The RL framework contains multiple agents with independent data processing capability~\cite{WJZRCH2020}. Each agent in the RL framework has a discrete set of environment states and a discrete set of agent actions. In each step, an agent chooses an action to perform from its action space, and changes its state accordingly. An agent learns how to choose the optimal action by continuously interacting with the system environment. Specifically, in each step, the agent observes the current state of the environment, and the systematic trial and error about state transition is indicated by a scalar reinforcement signal, i.e., reward. For each state-action pair, the long-term reward is indicated by an expected discounted reward, i.e., Q-value~\cite{CP1992}. After selecting an action, the agent updates the Q-table which records Q values of all state-action pairs, and obtains the optimal policy accordingly which is used to choose an action to maximize the long-term reward in the next step.

When the state-action space is small, Q-learning can works efficiently by maintaining look-up tables for the update of Q-values. In a large state-action space system, the Q-learning algorithm can be improved with deep neural network. We can estimate the Q-values by a deep neural network function approximator, known as DQN~\cite{MKSRVBGRFOPBSAKKWLAWR2015}, which addresses the sophisticate mappings between the states of UAVs and their corresponding Q-values based on a large amount of training data. When a DQN is well-trained for a UAV, given its current state as the input of the DQN, we can obtain the its Q-values on taking different actions from the outputs of the DQN.
\subsection{Algorithm for Valid Data Transmission Maximization}\label{RL trajectory subsection}
In this part, we describe the RL based algorithm that maximizes the average number of valid data transmissions. In each cycle, the UAVs design their own trajectories with a multi-agent deep RL based algorithm, the BS then performs radio resource management for the UAVs jointly as introduced in Section~\ref{Resource Sec}. To solve the UAV trajectory design problem with RL, we first formulate it into a Markov decision problem~\cite{SB1998}. The trajectory of the UAV impacts both the successful sensing probability and the successful transmission probability. We use a nested bi-level Markov chains to describe the sensing and transmission processes under the joint sensing and transmission protocol introduced in Section~\ref{protocol-sec}, where the outer Markov chain depicts the state transitions in UAV sensing, and the inner Markov chain depicts the state transitions in UAV transmission. The outer Markov chain contains two states: successful sensing and sensing failure, and the state transition takes place among different cycles. The inner Markov chain contains four states: successful transmission in U2N mode, successful transmission in U2U mode, successful transmission in U2D mode, and transmission failure, and the state transition takes place among different frames in a cycle.

\begin{figure}[t]
\centerline{\includegraphics[width=3in]{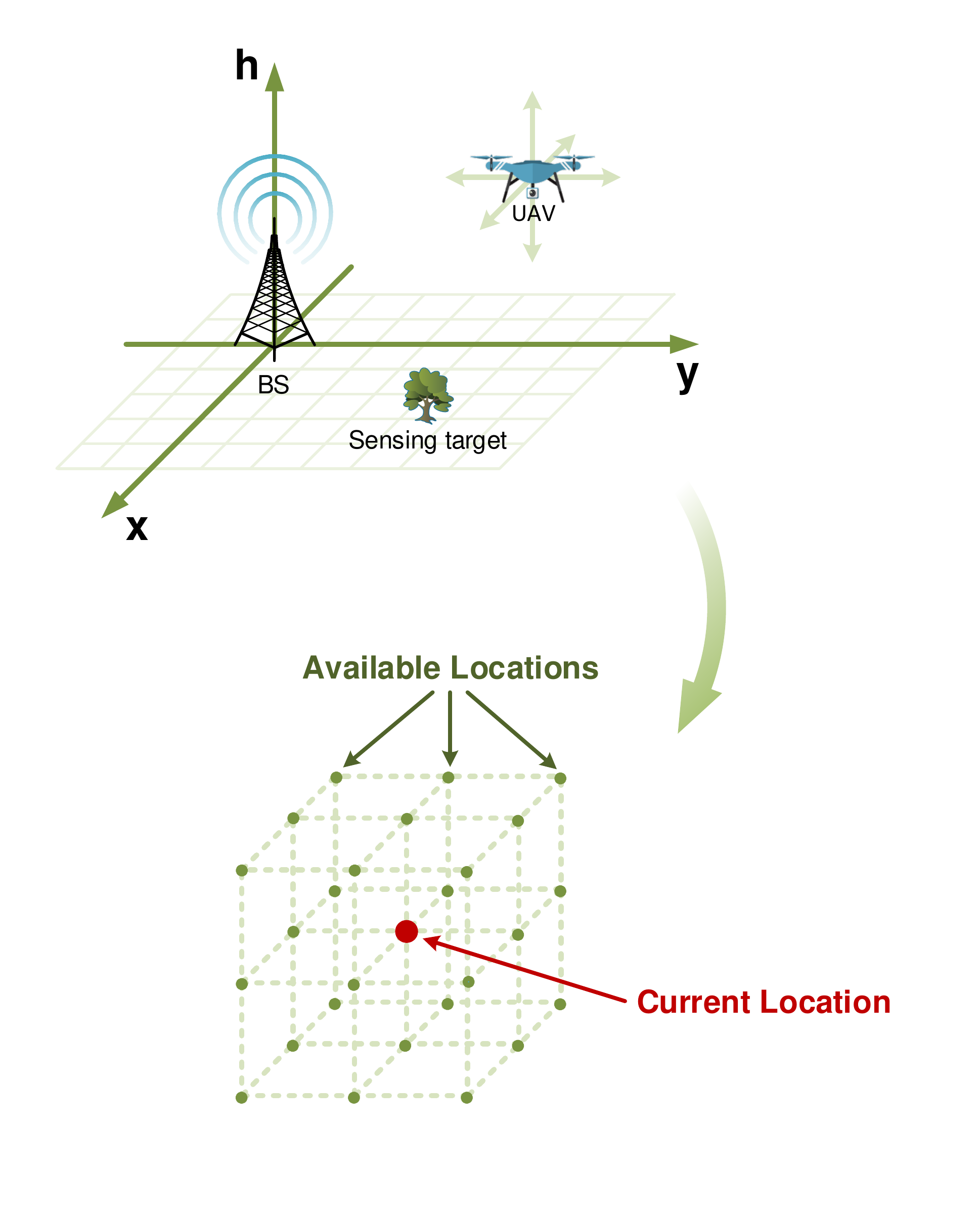}}
\caption{Available trajectories for a UAV in each cycle.}
\label{lattice}
\end{figure}

Since the UAV sensing and UAV transmission processes have been formulated as the state transitions in the nested bi-level Markov chains, the UAV trajectory design can be solved with RL algorithm. Under the RL framework, each UAV is regarded as an agent. The states and the actions of each agent are defined as the locations and the movements of the corresponding UAV, respectively. To describe the UAV trajectory with a finite set of actions, we divided the continuous space into a finite set of discrete spatial points as shown in Fig.~\ref{lattice}. Each UAV has 27 available locations in the next cycle in the 3-dimensional space, which corresponds to 27 different trajectories. We set the reward of an agent in one cycle as the expectation of its number of valid data transmissions. Since the trajectory in different cycles are connected, the agents have to consider the number of valid data transmissions in a sequence of cycles. Considering the timeliness requirements of sensing tasks, we also introduce a discounting valuation for the reward function of the agents in the future cycles. The discounted parameter is exponentially related to the time interval from the current cycle to the future one, and the utility of an agent in a cycle is defined as its total discounted rewards in the future. The trajectory design problem is then converted to maximizing the total expected discounted rewards of all agents by optimizing their state transition functions.

The Q value of an agent is set as the expected accumulated discounted rewards when it takes an action at a specific state, and the optimal Q-values of each agent can be obtained with an iterative algorithm. However, in the cellular Internet of UAVs, the state-action space of the trajectory design problem is too large to maintain look-up tables for the update of Q-values. Therefore, we bring in the DQN algorithm to maximize the Q values of the agents. In order to train the DQN, we utilize a separate target network to generate the target for training~\cite{MKSRVBGRFOPBSAKKWLAWR2015}. During the training of the Q values, we update its weight by minimizing a loss function related to the target network and the DQN. The value of the target network is updated to that of the DQN in every few cycles. The steps of the DQN-based UAV trajectory design algorithm is summarized as below.
\begin{itemize}
\item \textbf{Step 1:} Initialize the DQN with random weight, add a copy of the initial DQN as the target network;
\item \textbf{Step 2:} In each cycle, select the action that maximizes the Q value with a probability of $\epsilon$, select a random action with a probability of $1-\epsilon$;
\item \textbf{Step 3:} Perform the action and observe its reward and the next state;
\item \textbf{Step 4:} Store the sample into a replay memory;
\item \textbf{Step 5:} Sample a mini-batch from the replay memory;
\item \textbf{Step 6:} Perform a gradient descend step on the loss function with respect to the weight of the DQN using the mini-batch data set;
\item \textbf{Step 7:} Update the target network to the current DQN in every $r$ cycles;
\item \textbf{Step 8:} Go to the next cycle, and return to Step 2;
\end{itemize}

After the UAVs design their trajectories with RL algorithm, the BS performs radio resource management for them jointly. In the RL framework, the results of the radio resource management are the system environment of the agents. The subchannel and power control, which determines the rewards of the actions, can be solved as described in Section~\ref{Resource Sec}. Since the radio resource management contains both binary variables for subchannel allocation and continuous variables for power control, it can be proved to be NP-hard~\cite{M2012}. To solve this problem, the subchannel allocation and power control are decoupled into two subproblems, and solved by the BS iteratively until convergence. The subchannel allocation subproblem can be solved with the matching based algorithm introduced in Section~\ref{subcahnnel Sec}. The utility of the UAV is set as its transmission rate, and the utility of a subchannel is defined as the sum of the valid data transmissions over it. The power control subproblem can be solved with the difference of convex based algorithm as introduced in Section~\ref{power Sec}, and the weight coefficient for the UAV utility is a function of the size of data that needs to be transmitted.  

\subsection{Performance Analysis}
In the following, we present the performance of the proposed scheme. We assume that a total number of 5 UAVs and 2 terrestrial UEs are deployed in a cell. Each UAV has one sensing task and one transmission destination. The UAVs perform sensing and transmission as introduced in the joint sensing and transmission protocol for U2X communications. The UAV trajectory design, subchannel allocation, and power control are performed in the RL based mathematical framework as mentioned above. The maximum transmission power of a UAV is 10 dBm. The unit length of UAV trajectory in the deep RL algorithm is 25 m, and the transmission QoS threshold is set as 1 bit/s/Hz.

\begin{figure}[t]
\centerline{\includegraphics[width=3in]{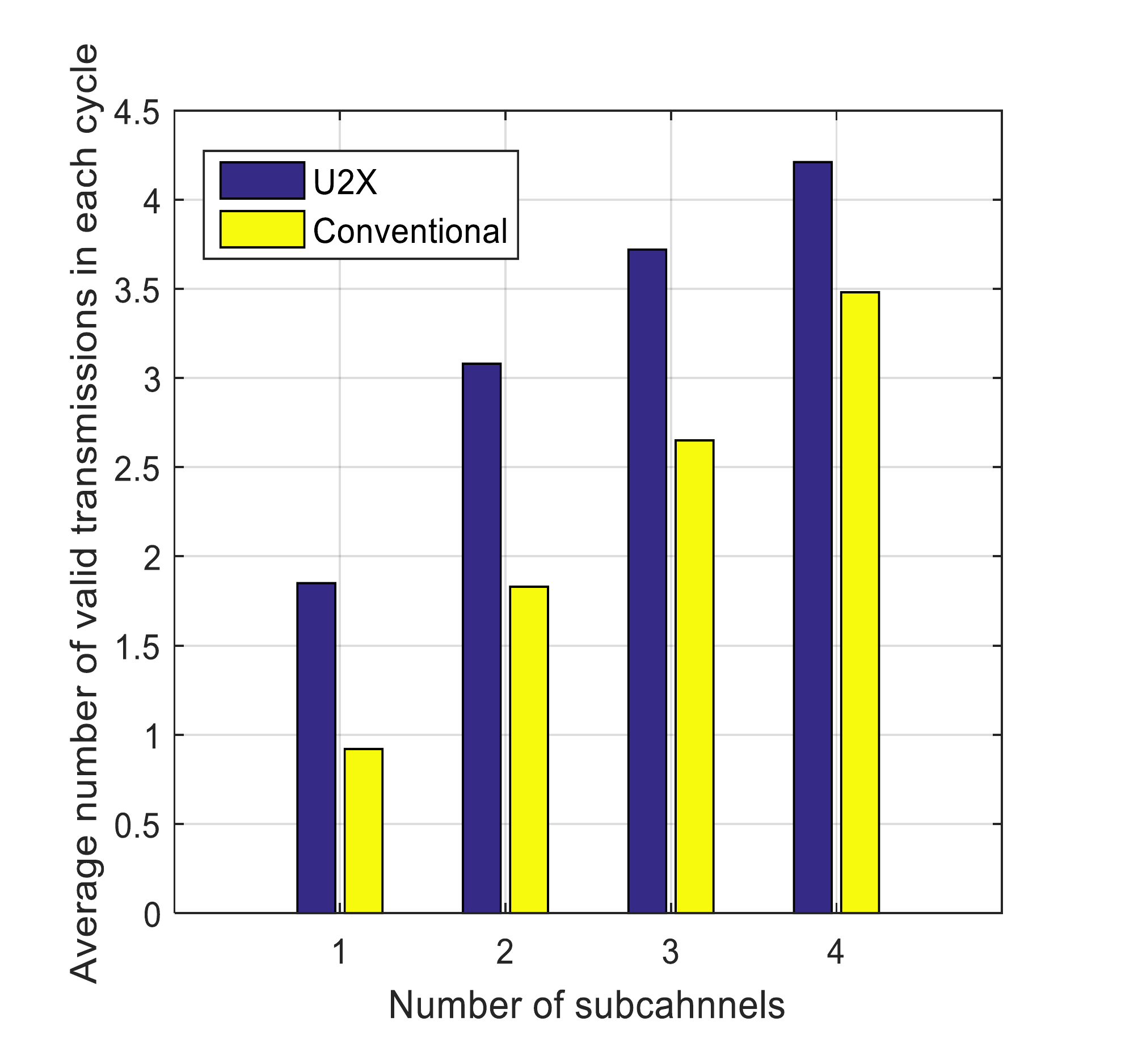}}
\caption{Simulation for average number of valid data transmissions.}
\label{simulation}
\end{figure}
Fig.~\ref{simulation} shows the average number of valid data transmissions in each cycle. We compare the proposed U2X communication framework to the conventional cellular one, in which all the UAVs first transmit the sensory data to the BS in U2N mode, and then the BS delivers the sensory data to the corresponding UEs. It is shown that the proposed U2X communications improve the number of valid data transmissions for over 30\% on average when compared to the conventional cellular mode due to the more flexible communication framework. The performance gap is even larger when the number of subchannels is small. The reason is that the U2U and U2D communications are capable to underlay the terrestrial cellular communication links, which enhance the spectrum efficiency of the network.

\begin{figure}[t]
\centerline{\includegraphics[width=3in]{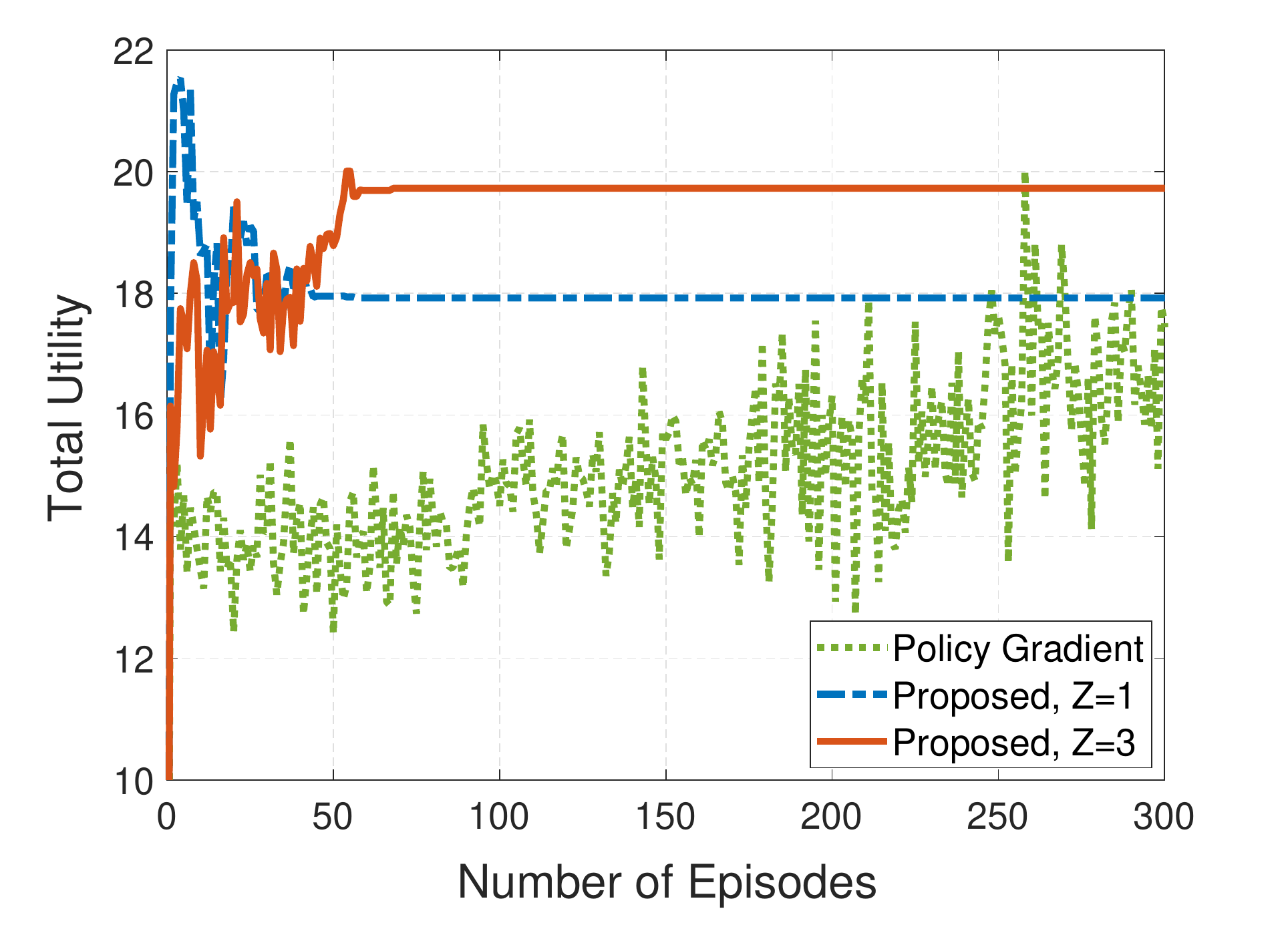}}
\caption{Simulation for network total utility.}
\label{simulation2}
\end{figure}
In Fig.~\ref{simulation2}, we present the total utility of the U2X communications with the proposed algorithm and a compared policy gradient algorithm, in which each UAV directly optimizes its parameterized control policy by a variant of gradient descent. Two different schemes of the proposed algorithm are given in this figure, with the number of hidden layers $Z$ in the DQN network being 1 and 3, respectively. The proposed algorithm converges within 70 episodes, which significantly outperforms the policy gradient algorithm. The total utility of the UAVs with the proposed algorithm can be over 15\% higher than that of the compared algorithm. A larger number of hidden layers can further improve the network utility by about 10\%, but it requires a larger number of episodes to converge.

\begin{figure}[t]
\centerline{\includegraphics[width=3in]{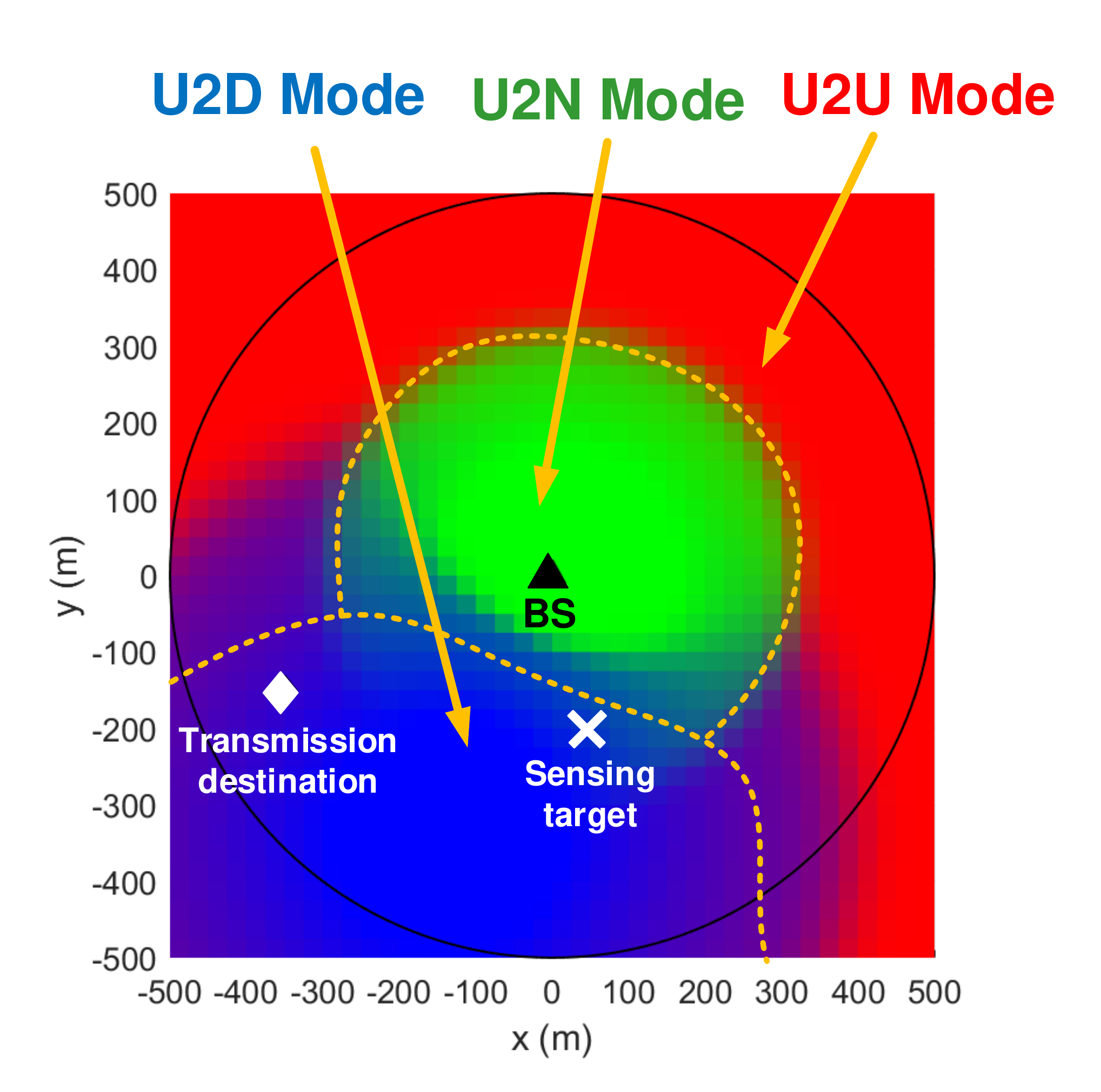}}
\caption{Simulation for transmission mode selection.}
\label{simulation3}
\end{figure}
Fig.~\ref{simulation3} illustrates the communication mode of a UAV, given the locations of its sensing target and transmission destination. The UAV adopts the U2D mode for data transmission when it is close to its transmission destination. Besides, as the UAV moves away from its transmission destination and close to the BS, it prefers to transmit in the U2N mode. When the UAV is far from its transmission destination as well as the BS, U2U mode is the optimal one for the UAV, i.e., it first transmits the sensory data to an adjacent UAV, and then the adjacent UAV relays the data to the transmission destination.

\begin{figure}[t]
    \centering
    \subfigure[]{
    \begin{minipage}{3in}\label{Simulation4-1}
    \includegraphics[width=3in]{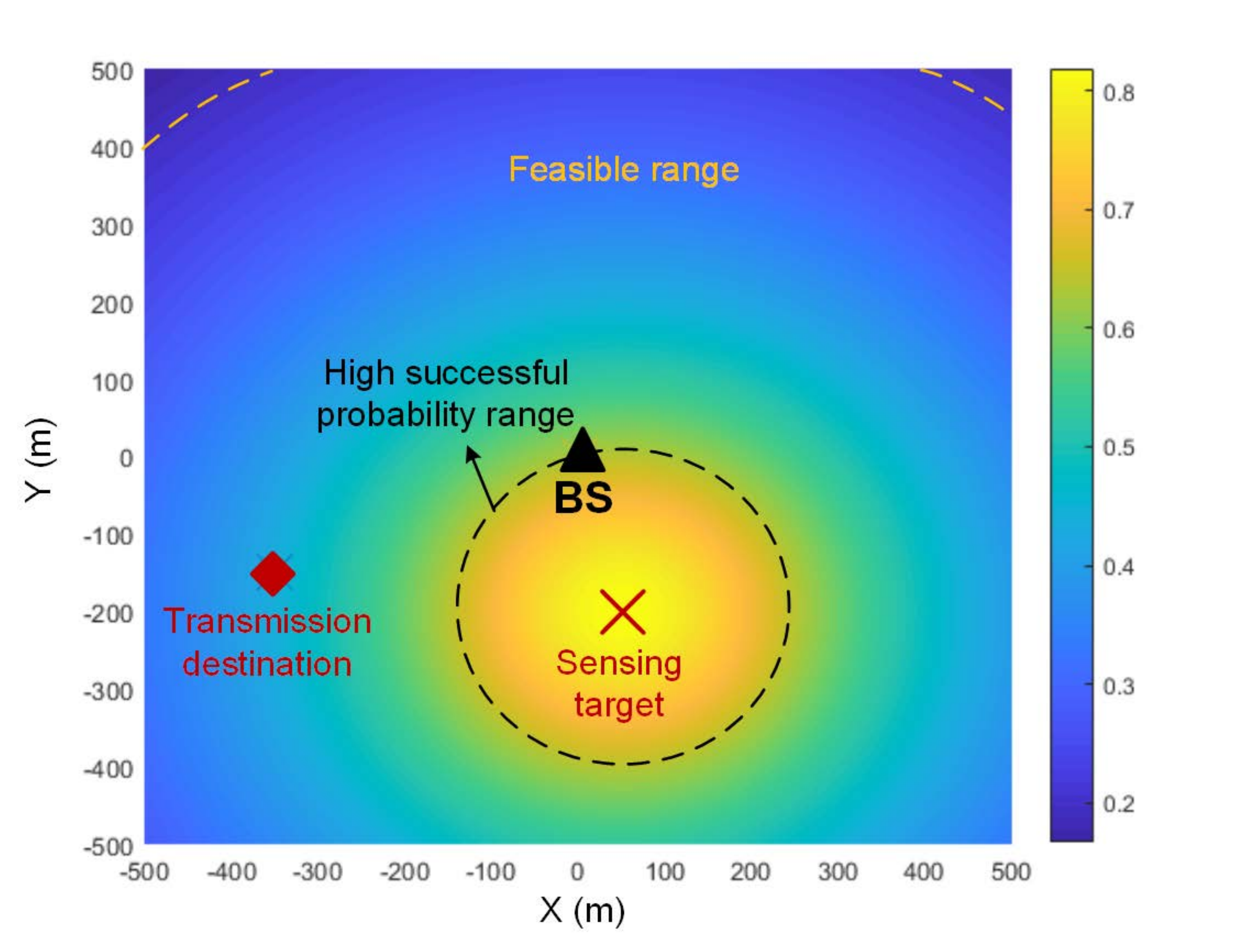}
    \end{minipage}
    }
    \subfigure[]{
    \begin{minipage}{3in}\label{Simulation4-2}
    \includegraphics[width=3in]{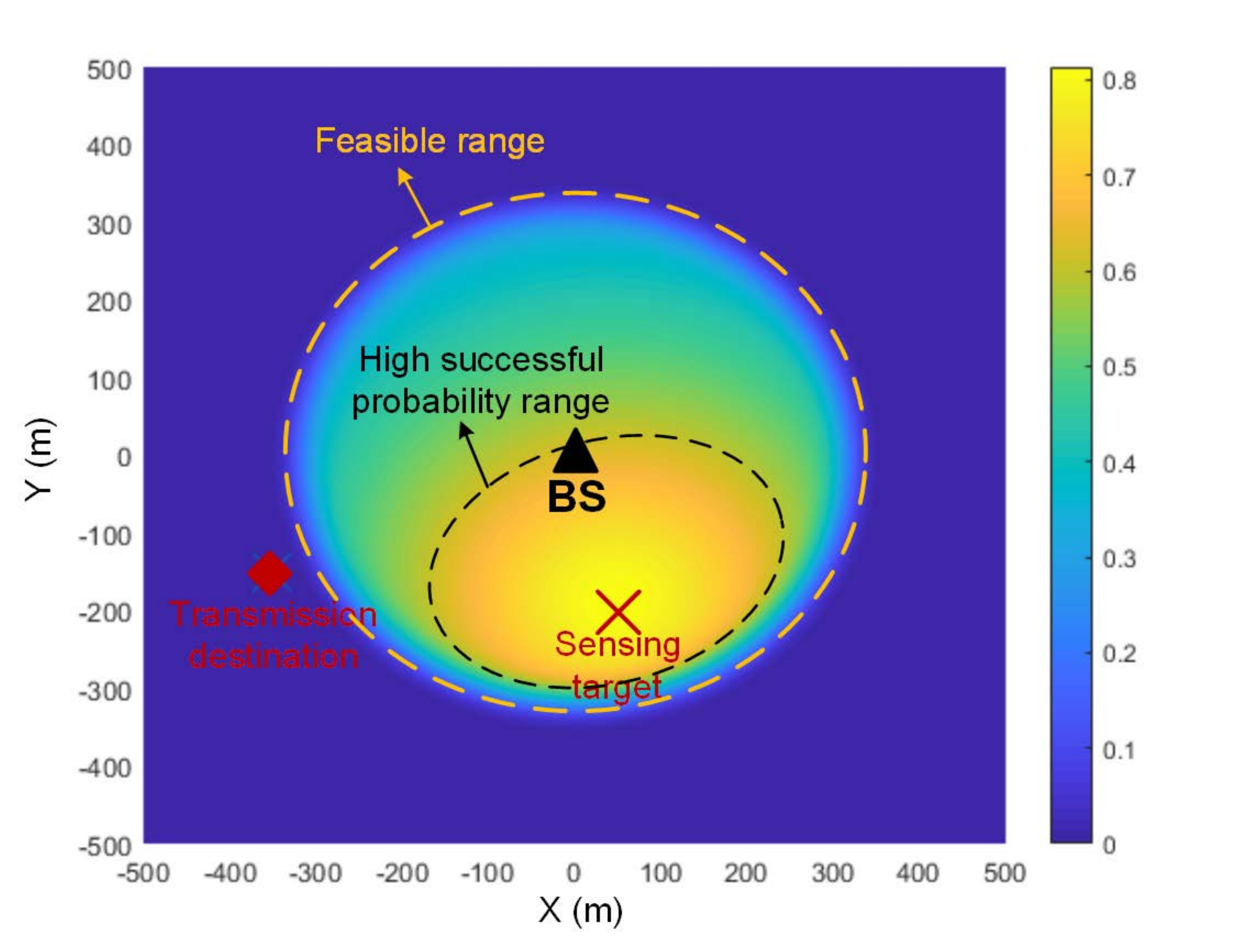}
    \end{minipage}
    }
    \caption{Simulation for successful sensing and transmission probability: a) U2X communications, b) Conventional cellular communications.}
    \label{Simulation4}
\end{figure}

Fig.~\ref{Simulation4} depicts the successful sensing and transmission probability of a UAV, given the locations of its sensing target and transmission destination. Fig~\ref{Simulation4-1} is the successful sensing and transmission probability with the proposed U2X communications. It can be shown that the sensing and transmission processes are feasible in most of the locations within the cell. The successful sensing and transmission probability is mostly affected by the distance between the UAV and the sensing target, which indicates that the successful sensing probability is the main impact factor, and the successful transmission probability keeps at a high level with U2X communications. In comparison, we given the successful sensing and transmission probability with only conventional cellular transmissions in Fig.~\ref{Simulation4-2}. In the compared scheme, the sensing and transmission can only be performed when the UAV is close to both the sensing target and the BS.
\section{Extensions with U2X Communications}\label{application sec}
In this section, we present three possible extensive scenarios for the U2X communications described above. Open problems and potential solutions are also discussed.
\subsection{UAV Cooperation with U2X Communications}
One promising application of the U2X communications is to achieve UAV cooperation in the cellular Internet of UAVs, in which multiple UAVs perform sensing and transmission of a task cooperatively~\cite{ZZDS2019'}. It is of great significance in performing tasks with large data rate from remote area, such as live video streaming~\cite{PAR2017}. To complete such an application, the UAVs that are remote from the transmission destination need to transmit their collected data in realtime with high data rate. Such an application presents challenges on the conventional cellular network for two reasons. First, it requires a large bandwidth to support the high rate upload for multiple UAVs simultaneously, which affects the transmission QoSs of the terrestrial UEs severely. Second, for the UAVs that are remote from the transmission destination, the data transmission requires large transmission power, which reduces the working lifetime of the UAVs. The U2X communication provides more flexible transmissions for the cooperative UAVs, thus providing the possibility to complete the above application with less spectrum and power consumptions~\cite{TMFKAIM2017}.

\begin{figure}[t]
\centerline{\includegraphics[width=3in]{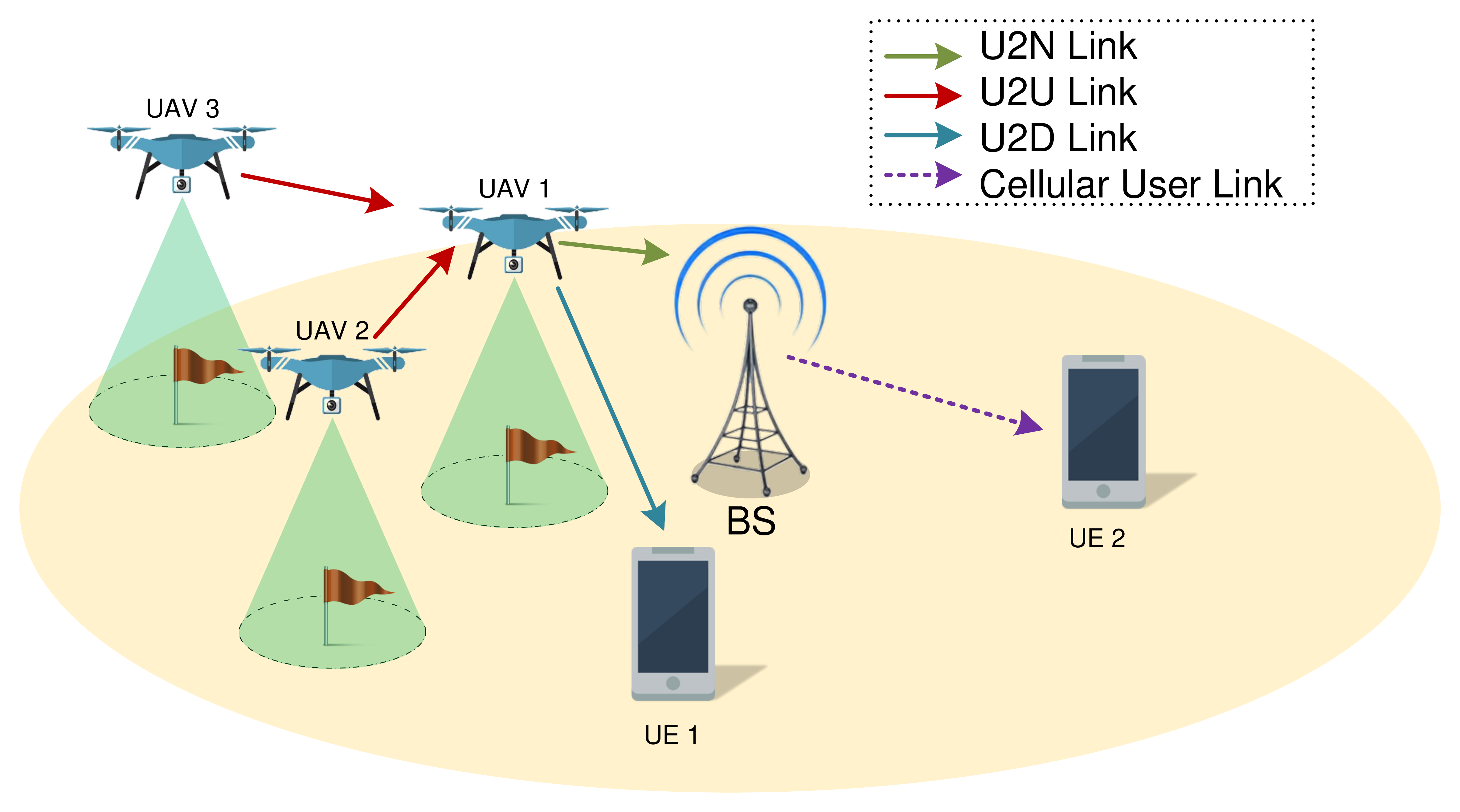}}
\caption{System model of UAV cooperation with U2X communications.}
\label{adhoc}
\end{figure}

Fig.~\ref{adhoc} is an example of the UAV cooperation with U2X communications. With the concept of U2X communications, the data sensed by different UAVs can be transmitted to the destination with multiple hops. The data sensed by different UAVs are transmitted to a UAV that has the best channel condition to the transmission destination in U2U mode. The UAV then transmit all the data of the cooperative UAVs to the BS in U2N mode, or transmits the data to the destination UE in U2D mode. In this way, the data transmission only overlays the bandwidth of one terrestrial UE, and the transmission power consumption is minimized.

Unlike the conventional multihop transmission networks, the topology of the cellular Internet of UAVs changes rapidly, and the density of the UAV nodes is usually lower than that of the terrestrial wireless sensor nodes. Therefore, the routing protocol of the cooperative transmission needs to be further studied. Recently, some studies start the design of multihop transmission routing protocols for the cellular Internet of UAVs~\cite{ZSW2018}. However, most of the existing protocols are heuristic, and few of them considers the onboard energy and buffer of the UAVs, which significantly affects the performance of the UAVs in practical systems. A protocol that jointly considers the trajectories and physical constraints of the UAVs needs to be designed in the future works.
\subsection{Mobile Edge Computing with U2X Communications}
MEC with U2X communications provides additional computing capabilities for the UEs that are remote from the BS~\cite{JSK2018}. It is especially helpful for the sensing tasks performed in a wide range with data processing requirements, such as precision agriculture. In a conventional cellular network, the UAV needs to upload the collected data to the BS. The BS then performs data computation, and transmits the processed data to all the requiring UEs. Such a framework poses huge computation and transmission pressure on the network infrastructure, which not only causes long computation delay at the BS, but also leads to low transmission QoS for the cell edge UEs~\cite{TFMKOM2019}. To tackle this problem, we utilize MEC to offload the data computation in the cellular Internet of UAVs.

\begin{figure}[t]
\centerline{\includegraphics[width=3in]{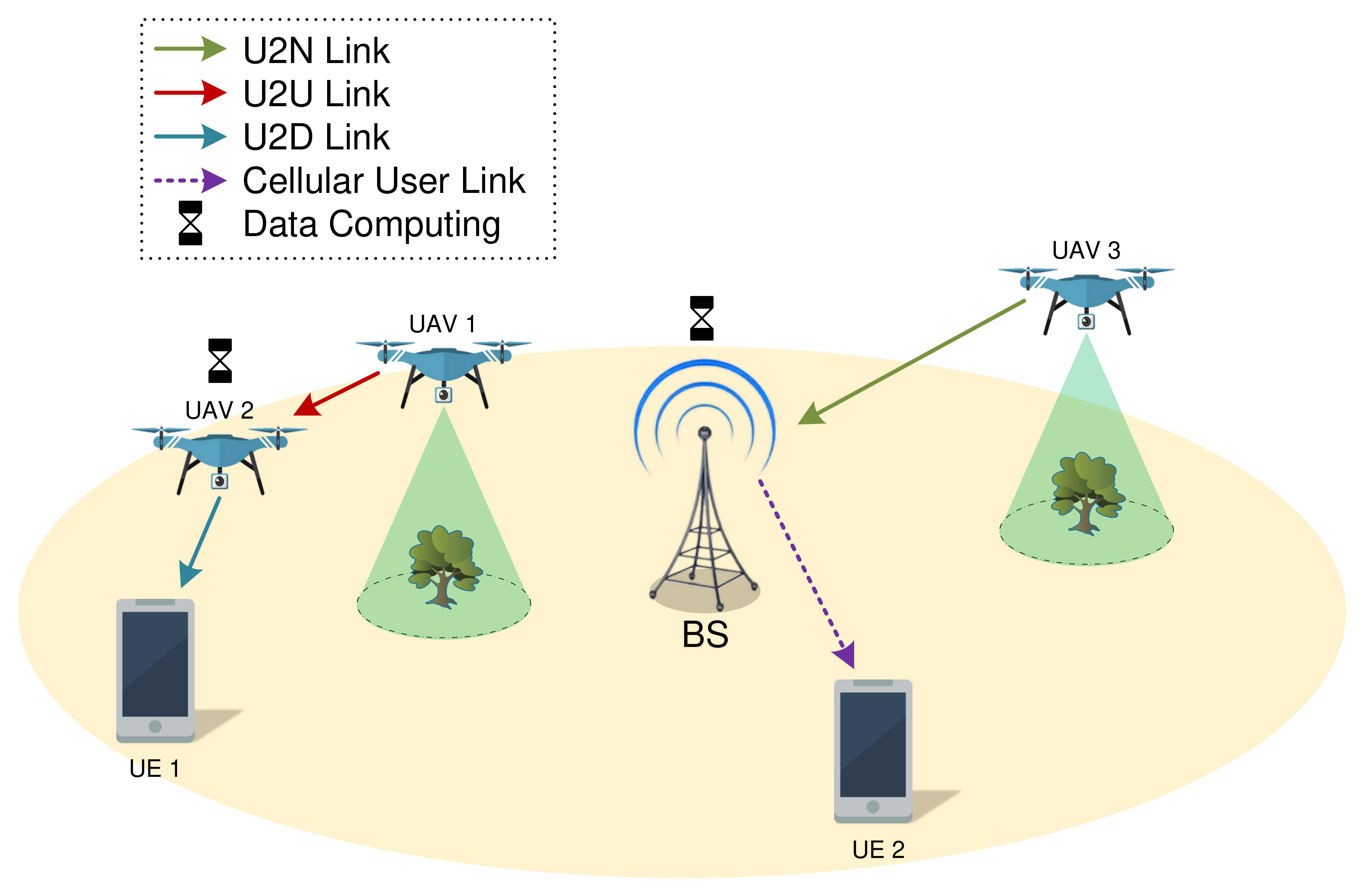}}
\caption{System model of MEC with U2X communications.}
\label{offloading}
\end{figure}

We consider a cellular Internet of UAVs as shown in Fig.~\ref{offloading}, where the UAVs collect sensory data that needs to be processed before transmitting to the destination UEs. To reduce the computation workload of the BS, some of the sensory data can be offloaded to the cell edge UAVs with computing capability in U2U mode. After data processing, the cell edge UAVs transmit the data to the destination UEs directly in U2D mode. The MEC with U2X communications reduces the computation pressure of the BS, and enhances the transmission QoS to the cell edge UEs significantly by reducing the transmission distance. There are still some open problems to be solved for the utilization of MEC with U2X communications. First, the computation resource for different tasks should be allocated properly. To be specific, which task is computed by the BS, and which task is computed by the UAVs. Second, the energy consumption of the MEC UAVs should be considered since their onboard batteries are limited.

\subsection{Non-Orthogonal Multiple Access for U2X Communications}
The rapid development of cellular Internet of UAVs poses challenges on the conventional orthogonal multiple access systems, where UAVs have to share limited radio resources in an orthogonal manner. The orthogonal multiple access scheme suffers from serious congestion problems when a large number of UAVs intend to transmit their sensory data concurrently. To tackle the challenges of access collision reduction and massive connectivity, NOMA scheme has been introduced as a potential solution, which allows the UAVs to access the radio resources non-orthogonally.

NOMA can be utilized for U2X communications where multiple UAVs are allowed to access the channel non-orthogonally by either power domain~\cite{ZDSL2017} or code domain~\cite{DSLL2019} multiplexing. Multiple UAVs with different communication modes can transmit concurrently on the same channel to improve the spectrum efficiency and reduce the transmission latency. To make the NOMA scheme more practical, the multi-user detection technique such as successive interference cancellation, can be applied at the transmission destinations for decoding, to cope with the co-channel interference caused by spectrum sharing. Due to the high mobility and 3-dimensional space properties of the cellular UAVs, the design of NOMA-based cellular U2X communications becomes different from the traditional NOMA system in many aspects, including spectrum management, power control, signaling control, and needs to be further studied in the future.

\section{Conclusions}\label{conclusion sec}
In this article, we have considered the utilization of full dimension U2X communications in the cellular Internet of UAVs, which enables the UAVs to support various sensing applications with high transmission rate. We have proposed a basic model of the cellular Internet of UAVs, and have introduced the three transmission modes, i.e., U2N mode, U2U mode, and U2D mode for the U2X communications. Based on the basic model, we have raised three key technologies that need to be studied to improve the transmission QoS performance, including joint sensing and transmission protocol for U2X communications, UAV trajectory design, and radio resource management. We have proposed a RL based mathematical framework to maximize the number of valid data transmissions with U2X communications. The extensions and open problems of U2X communications have also been discussed, such as UAV cooperation, MEC, and NOMA with U2X communications.


\begin{thebibliography}{15}
\bibitem{DOCOMO}
``White paper: 5G evolution and 6G," \emph{NTT DOCOMO Inc.,} Jan. 2020.
\bibitem{ZXMXDLKF2019}
Z. Zhang, Y. Xiao, Z. Ma, M. Xiao, Z. Ding, X. Lei, G. K. Karagiannidis, and P. Fan, ``6G wireless networks: vision, requirements, architecture, and key technologies," \emph{IEEE Veh. Technol. Mag.,} vol.~14, no.~3, pp.~28-41, Sep.~2019.
\bibitem{ZFWDWW2019}
B. Zong, C. Fan, X. Wang, X. Duan, B. Wang, and J. Wang, ``6G technologies: key drivers, core requirements, system architectures, and enabling technologies," \emph{IEEE Veh. Technol. Mag.,} vol.~14, no.~3, pp.~18-27, Sep.~2019.
\bibitem{FTNKM2016}
Z. M. Fadlullah, D. Takaishi, H. Nishiyama, N. Kato, and R. Miura, ``A dynamic trajectory control algorithm for improving the communication throughput and delay in UAV-aided networks," \emph{IEEE Network,} vol.~30, no.~1, pp.~100-105, Jan.~2016.
\bibitem{ZZHZXWWZ2019}
L. Zhang, H. Zhao, S. Hou, Z. Zhao, H. Xu, X. Wu, Q. Wu, and R. Zhang, ``A survey on 5G millimeter wave communications for UAV-assisted wireless networks," \emph{IEEE Access,} vol.~7, pp.~117460-117504, Jul.~2019.
\bibitem{UAVmarket}
D. Joshi, \emph{Commercial unmanned aerial vehicle (UAV) market analysis-industry trends, companies and what you should know.} Accessed on Aug. 2017, available:
http://www.businessinsider.com/commercial-uav-market-analysis-2017-8.
\bibitem{ZSH2019}
H. Zhang, L. Song, and Z. Han, \emph{Unmanned Aerial Vehicle Applications over Cellular Networks for 5G and Beyond}. New York: Springer, 2020.
\bibitem{KNKOM2019}
Y. Kawamoto, H. Nishiyama, N. Kato, F. Ono, and R. Miura, ``Toward future unmanned aerial vehicle networks: architecture, resource allocation and field experiments," \emph{IEEE Wireless Commun.,} vol.~26, no.~1, pp.~94-99, Feb.~2019.
\bibitem{KGM2017}
T. Kersnovski, F. Gonzalez, and K. Morton, ``A UAV system for autonomous target detection and gas sensing," in \emph{Proc. IEEE Aerospace Conf.,} Big Sky, MT, Mar.~2017.
\bibitem{AMCG2017}
B. H. Y. Alsalam, K. Morton, D. Campbell, and F. Gonzalez, ``Autonomous UAV with vision based on-board decision making for remote sensing and precision agriculture," in \emph{Proc. IEEE Aerospace Conf.,} Big Sky, MT, Mar.~2017.
\bibitem{ZZHBS2017}
S. Zhang, H. Zhang, Q. He, K. Bian, and L. Song, ``Joint trajectory and power optimization for UAV relay networks," \emph{IEEE Commun. Lett.,} vol.~22, no.~1, pp.~161-164, Jan.~2018.
\bibitem{XZLNWG2020}
Z. Xiong, Y. Zhang, N. C. Luong, D. Niyato, P. Wang, and N. Guizani, ``The best of both worlds: a general architecture for data management in blockchain-enabled Internet-of-Things," \emph{IEEE Network,} vol.~34, no.~1, pp.~166-173, Jan. 2020.
\bibitem{TFKOM2018}
F. Tang, Z. M. Fadlullah, N. Kato, F. Ono, and R. Miura, ``AC-POCA: anticoordination game based partially overlapping channels assignment in combined UAV and D2D-Based networks," \emph{IEEE Trans. Veh. Technol.,} vol.~67, no.~2, pp.~1672-1683, Feb.~2018.
\bibitem{ZZDS2019}
S. Zhang, H. Zhang, B. Di, and L. Song, ``Cellular UAV-to-X communications: design and optimization for multi-UAV networks," \emph{IEEE Trans. Wireless Commun.,} vol.~18, no.~2, pp.~1346-1359, Jan.~2019.


\bibitem{ZSHP2019}
H. Zhang, L. Song, Z. Han, and H. V. Poor, ``Cooperation techniques for a cellular Internet of unmanned aerial vehicles," \emph{IEEE Wireless Commun.,} vol.~26, no.~5, pp.~167-173, Oct.~2019.
\bibitem{GLTKA2020}
G. Gui, M. Liu, F. Tang, N. Kato, and F. Adachi, ``6G: Opening new horizons for integration of comfort, security and intelligence," \emph{IEEE Wireless Commun. Mag. (Early Access),} Mar. 2020.
\bibitem{KMTKL2020}
N. Kato, B. Mao, F. Tang, Y. Kawamoto, and J. Liu, ``Ten challenges in advancing machine learning technologies towards 6G," \emph{IEEE Wireless Commun. Mag. (Early Access),} Mar. 2020.
\bibitem{CHSPFZZ2017}
S. Chen, J. Hu, Y. Shi, Y. Peng, J. Fang, R. Zhao, and L. Zhao, ``Vehicle-to-everything (v2x) services supported by LTE-based systems and 5G," \emph{IEEE Commun. Standard Mag.,} vol.~1, no.~2, pp.~70-76, Jun.~2017.
\bibitem{DSLH2017}
B. Di, L. Song, Y. Li, and Z. Han, ``V2X meets NOMA: non-orthogonal multiple access for 5G-enabled vehicular networks," \emph{IEEE Wireless Commun.,} vol.~24, no.~6, pp.~14-21, Dec.~2017.
\bibitem{JZG2019}
H. Jiang, Z. Zhang, and G. Gui, ``Three-dimensional non-stationary wideband geometry-based UAV channel model for A2G communication environments," \emph{IEEE Access,} vol.~7, pp.~26116-26122, Feb.~2019.

\bibitem{ZZDS2019'}
S. Zhang, H. Zhang, B. Di, and L. Song, ``Cellular cooperative unmanned aerial vehicle networks with sense-and-send protocol," \emph{IEEE Internet Things J.,} vol.~18, no.~2, pp.~1346-1359, Jan.~2019.
\bibitem{ZYZS2019}
S. Zhang, J. Yang, H. Zhang, and L. Song, ``Dual trajectory optimization for a cooperative Internet of UAVs," \emph{IEEE Commun. Lett.,} vol.~23, no.~6, pp.~1093-1096, Jun.~2019.

\bibitem{3GPPR12}
3GPP TS 36.777, ``Enhanced LTE support for aerial vehicles," Release~15, Dec.~2017.

\bibitem{ZLS2017}
H. Zhang, Y. Liao, and L. Song, ``D2D-U: Device-to-device communications in unlicensed bands for 5G system," \emph{IEEE Trans. Wireless Commun.,} vol.~16, no.~6, pp.~3507-3519, Jun.~2017.

\bibitem{LSHL2015}
Y. Liao, L. Song, Z. Han, and Y. Li, ``Full duplex cognitive radio: a new design paradigm for enhancing spectrum usage," \emph{IEEE Commun. Mag.,} vol.~53, no.~5, pp.~138-145, May~2015.

\bibitem{TKNKOM2018}
Y. Takahashi, Y. Kawamoto, H. Nishiyama, N. Kato, F. Ono, and R. Miura, ``A novel radio resource optimization method for relay-based unmanned aerial vehicles," \emph{IEEE Trans. Wireless Commun.,} vol.~17, no.~11, pp.~7352-7363, Nov.~2018.
\bibitem{TKKL2019}
F. Tang, Y. Kawamoto, N. Kato, and J. Liu, ``Future intelligent and secure vehicular network toward 6G: machine-learning approaches," \emph{Proc. IEEE,} vol.~108, no.~2, pp.~292-307, Feb.~2020.
\bibitem{XZNDWW2019}
Z. Xiong, Y. Zhang, D. Niyato, R. Deng, P. Wang, and L. Wang, ``Deep reinforcement learning for mobile 5G and beyond: fundamentals, applications and challenges", \emph{IEEE Veh. Technol. Mag.,} vol.~14, no.~2, pp.~44-52, Apr.~2019.
\bibitem{ZDSL2017}
S. Zhang, B. Di, L. Song, and Y. Li, ``Sub-Channel and power allocation for non-orthogonal multiple access relay networks with amplify-and-forward protocol," \emph{IEEE Trans. Wireless Commun.,} vol.~16, no.~4, pp.~2249-2261, Apr.~2017.
\bibitem{TKNKOM2018'}
Y. Takahashi, Y. Kawamoto, H. Nishiyama, N. Kato, F. Ono, and R. Miura, ``Virtual cell based resource allocation for efficient frequency utilization in unmanned aircraft systems," \emph{IEEE Trans. Veh. Technol.,} vol.~67, no.~4, pp.~3495-3504, Apr.~2018.
\bibitem{HZS2019}
J. Hu, H. Zhang, and L. Song, ``Reinforcement learning for decentralized trajectory design in cellular UAV networks with sense-and-send protocol," \emph{IEEE Internet Things J.,} vol.~6, no.~4, pp.~6177-6189, Aug.~2019.
\bibitem{WJZRCH2020}
J. Wang, C. Jiang, H. Zhang, Y. Ren, K. C. Chen, and L. Hanzo, ``Thirty years of machine learning: the road to pareto-optimal wireless networks," \emph{IEEE Commun. Surveys Tuts. (Early Access),} Jan.~2020.

\bibitem{CP1992}
C. Watkins and P. Dayan, ``Q-learning," \emph{Mach. Learn.,} vol.~8, pp.~279-292,~1992.
\bibitem{MKSRVBGRFOPBSAKKWLAWR2015}
V. Mnih, K. Kavukcuoglu, D. Silver, A. A. Rusu, J. Veness, M. G. Bellemare, A. Graves, M. Riedmiller, A. K. Fidjeland,
G. Ostrovski, S. Petersen, C. Beattie, A. Sadik, I. Antonoglou, H. King, D. Kumaran, D. Wierstra, S. Legg,
D. H. I. Antonoglou, D. Wierstra, and M. A. Riedmiller, ``Human-level control through deep reinforcement learning,"
\emph{Nature}, vol.~518, no.~7540, pp.~529-533, Feb.~2015.
\bibitem{SB1998}
R. S. Sutton and A. G. Barto, \emph{Reinforcement learning: an introduction}. Cambridge, MA: MIT press, 1998.
\bibitem{M2012}
M. Sipser, \emph{Introduction to the theory of computation}. 3rd ed., Boston, MA: Cengage Learning, pp.~225-277, 2012.

\bibitem{PAR2017}
L. R. Pinto, L. Almeida, and A. Rowe, ``Demo abstract: video streaming in multi-hop aerial networks," in \emph{Proc. IEEE IPSN,} Pittsburgh, PA, Apr.~2017.
\bibitem{TMFKAIM2017}
F. Tang, B. Mao, Z. Fadlullah, N. Kato, O. Akashi, T. Inoue, and K. Mizutani, ``On removing routing protocol from future wireless networks: a real-time deep learning approach for intelligent traffic control," \emph{IEEE Wireless Commun.,} vol.~25, no.~1, pp.~154-160, Feb.~2018.
\bibitem{ZSW2018}
Z. Zheng, A. K. Sangaiah, and T. Wang, ``Adaptive communication protocols in flying ad hoc network," \emph{IEEE Commun. Mag.,} vol.~56, no.~1, pp.~136-142, Jan.~2018.
\bibitem{JSK2018}
S. Jeong, O. Simeone, and J. Kang, ``Mobile edge computing via a UAV-mounted cloudlet: optimization of bit allocation and path planning," \emph{IEEE Trans. Veh. Technol.,} vol.~67, no.~3, pp.~2049-2063, Mar.~2018.
\bibitem{TFMKOM2019}
F. Tang, Z. Fadlullah, B. Mao, N. Kato, F. Ono, and R. Miura, ``On a novel adaptive UAV-mounted cloudlet-aided recommendation system for LBSNs," \emph{IEEE Trans. Emerging Topics Comp.,} vol.~7, no.~4, pp.~565-577, Oct.~2019.
\bibitem{DSLL2019}
B. Di, L. Song, Y. Li, and G. Y. Li, ``TCM-NOMA: joint multi-user codeword design and detection in trellis-coded modulation-based NOMA for beyond 5G," \emph{IEEE J. Sel. Topics Signal Process,} vol.~13, no.~3, pp.~766-780, Jun. 2019.

\end{thebibliography}
\end{document}